\documentclass[10pt,conference]{IEEEtran}

\usepackage[english]{babel} 
\usepackage{amsmath}                
\usepackage{amsfonts}               
\usepackage{amssymb}                
\usepackage{amsopn}                 
\allowdisplaybreaks[3] 
\usepackage{bbm}                    
\usepackage{mathrsfs}               
\usepackage{calc}                   
\usepackage[dvips]{graphicx}        
\usepackage{epsfig}                
\usepackage{psfrag}                 
\usepackage[dvips]{color}           
\usepackage{fancyhdr}               
\usepackage{verbatim}               
\usepackage{exscale}                
\usepackage{research5}
\newtheorem{theorem}{Theorem}

\newcommand{\PiG}{\Pi_{\textnormal{G}}}
\newcommand{\I}{\mathcal{I}}
\renewcommand{\I}{I}
\newcommand{\h}{h}
\newcommand{\C}{C}

\title{Gaussian Fading is the Worst Fading}

\author{\authorblockN{Tobias Koch ~~~ Amos Lapidoth}
\authorblockA{Signal and Information Processing Laboratory\\ETH Zurich\\
CH-8092 Zurich, Switzerland\\
Email: \{tkoch, lapidoth\}@isi.ee.ethz.ch}}

\begin{document}

\maketitle

\begin{abstract}
  The capacity of peak-power limited, single-antenna, non-coherent,
  flat-fading channels with memory is considered. The emphasis is on
  the capacity pre-log, i.e., on the limiting ratio of channel capacity to
  the logarithm of the signal-to-noise ratio (SNR), as the SNR tends
  to infinity. It is shown that, among all stationary \& ergodic
  fading processes of a given spectral distribution function whose law
  has no mass point at zero, the Gaussian process gives rise to the
  smallest pre-log.
\end{abstract}

\section{Introduction}
\label{sec:intro}

The subject of this paper is the capacity of peak-power limited,
single-antenna, discrete-time, non-coherent, flat-fading channels with
memory. The transmitter and receiver are both
aware of the law of the fading process, but not of its realization.
Our focus is on the capacity in the high signal-to-noise ratio (SNR)
regime. Specifically, we study the capacity pre-log, which is defined
as the limiting ratio of channel capacity to the logarithm of the SNR, as the
SNR tends to infinity. We show that of all stationary \& ergodic
fading processes of a given spectral distribution function and having
no mass point at zero, the Gaussian process yields the smallest
pre-log. To state this result precisely we begin with a description of
the channel model.

\subsection{Channel Model}

We consider a single-antenna flat-fading channel with memory where the
time-$k$ channel output $Y_k \in \Complex$ corresponding to the
time-$k$ channel input $x_k \in \Complex$ is given by
\begin{equation}
  \label{eq:channel}
  Y_k = H_k x_k + Z_k.
\end{equation}
Here $\Complex$ denotes the complex field, and the random processes
$\{Z_k\}$ and $\{H_k\}$ take value in $\Complex$ and model the
additive and multiplicative noises, respectively.  It is assumed that
these processes are statistically independent and of a joint law that
does not depend on the input sequence $\{x_k\}$. 

The additive noise $\{Z_k\}$ is a sequence of independent and
identically distributed (IID) zero-mean, variance-$\sigma^2$,
circularly-symmetric, complex Gaussian random variables. The
multiplicative noise (``fading'') $\{H_k\}$ is a mean-$d$,
unit-variance, stationary \& ergodic stochastic process with spectral
distribution function $F(\lambda)$, $-1/2 \leq \lambda\leq 1/2$, i.e.,
\begin{equation}
  \E{(H_{k+m}-d)\conj{(H_k-d)}} = 
  \int_{-1/2}^{1/2} e^{\ii 2 \pi m \lambda}\d F(\lambda)
\end{equation}
where $\ii = \sqrt{-1}$, and where $\conj{A}$ denotes the complex
conjugate of $A$.

\subsection{The Pre-Log}

The capacity of our channel under a peak-power constraint
$\const{A}^2$ on the input is given by
\begin{equation}
  \C(\SNR) = 
  \lim_{n \to \infty}\frac{1}{n} \sup \I(X_1^n;Y_1^n)
\end{equation}
where the SNR is defined as
\begin{equation}
  \SNR \triangleq \frac{\const{A}^2}{\sigma^2};
\end{equation}
$A_1^n$ denotes the sequence $A_1,\ldots,A_n$; and where the
maximization is over all joint distributions on $X_1,\ldots,X_n$ satisfying with
probability one
\begin{equation}
  \label{eq:power}
  |X_k|^2 \leq \const{A}^2, \qquad k=1,\ldots,n.
\end{equation}
The capacity \emph{pre-log} is now defined by 
\begin{equation}
  \Pi \triangleq \varlimsup_{\SNR \to \infty}
  \frac{\C(\SNR)}{\log \SNR}.
\end{equation}

For \emph{Gaussian fading}, i.e., when $\{H_{k}-d\}$ is a
circularly-symmetric, complex Gaussian
process, the pre-log $\PiG$ is given by the Lebesgue measure of the
set of harmonics where the derivative of the spectral distribution
function is zero, i.e., \cite{lapidoth05}, \cite{lapidoth03_2}
\begin{equation}
  \label{eq:gauss}
  \PiG = \mu\left(\left\{\lambda:F'(\lambda)=0\right\}\right)
\end{equation}
where $\mu(\cdot)$ denotes the Lebesgue measure on the interval
$[-1/2,1/2]$, and where $F'(\cdot)$ denotes the derivative of $F(\cdot)$.
(Here the subscript ``G'' stands for ``Gaussian''.)

This result indicates that if the fading process is Gaussian and
band-limited, then the corresponding channel capacity grows
logarithmically in the SNR. Note that otherwise the capacity can
increase with the SNR in various ways. For instance, in
\cite{lapidothmoser03_3} fading channels are studied that result in a
capacity which increases double-logarithmically with the SNR, and in
\cite{lapidoth05} spectral distribution functions are presented for
which capacity grows as a fractional power of the logarithm of the
SNR.

\subsection{The Main Result}
In this paper we show that the Gaussian fading has the lowest pre-log
among all fading processes having a given spectral distribution
function and having no mass point at zero
\begin{equation}
  \label{eq:condition}
  \Prv{H_k=0}=0, \qquad k\in\Integers.
\end{equation}
Thus, if the stationary \& ergodic process $\{H_{k}\}$ satisfies \eqref{eq:condition} and is of spectral
distribution function $F(\cdot)$ then
\begin{align}
  \Pi & \geq \PiG \nonumber\\
  & = \mu\left(\left\{\lambda:F'(\lambda)=0\right\}\right).
\end{align}
This is made precise in the following theorem.
\begin{theorem}
  \label{thm:prelog}
  Consider a mean-$d$, unit-variance, stationary \& ergodic fading
  process $\{H_k\}$ with spectral distribution function $F(\cdot)$ and
  satisfying
  \begin{equation}
    \Prv{H_k=0} = 0, \qquad k \in \Integers.
  \end{equation}
  Then, the corresponding capacity pre-log $\Pi$ is lower bounded by
  \begin{equation}
    \Pi \geq \mu\left(\left\{\lambda:F'(\lambda)=0\right\}\right).
  \end{equation}
\end{theorem}

To prove this theorem we propose in the next section a lower bound on
channel capacity and proceed in Section \ref{sec:prelog} to analyze
its asymptotic growth as the SNR tends to infinity.



\section{A Capacity Lower Bound}
\label{sec:lb}

To derive a lower bound on the capacity we consider inputs
$\{X_k\}$ that are IID, zero-mean, circularly-symmetric, and for which
$|X_k|^2$ is uniformly distributed on the interval
$\left[0,\const{A}^2\right]$. Our derivation is based on the lower bound
\begin{equation}
  \label{eq:1}
  \I(X_1^n;Y_1^n) \geq \I(X_1^n;Y_1^n|H_1^n) - \I(H_1^n;Y_1^n|X_1^n)
\end{equation}
which follows from the chain rule
\begin{IEEEeqnarray}{lCl}
\IEEEeqnarraymulticol{3}{l}{\I(X_1^n;Y_1^n)}\nonumber\\
\quad & = & \I(X_1^n,H_1^n;Y_1^n) -
\I(H_1^n;Y_1^n|X_1^n)\nonumber\\
& = & \I(H_1^n;Y_1^n) + \I(X_1^n;Y_1^n|H_1^n) - \I(H_1^n;Y_1^n|X_1^n) \IEEEeqnarraynumspace
\end{IEEEeqnarray}
and the non-negativity of mutual information.

We first study the first term on the right-hand side (RHS) of
\eqref{eq:1}. Making use of the stationarity of the channel and of
the fact that the inputs are IID we have
\begin{equation}
  \label{eq:coherent2}
  \lim_{n\to \infty}\frac{1}{n} \I(X_1^n;Y_1^n|H_1^n) = \I(X_1;Y_1|H_1). 
\end{equation}
We now lower bound the RHS of \eqref{eq:coherent2} as follows. For any
fixed $\Gamma > 0$
\begin{IEEEeqnarray}{lCl}
  \IEEEeqnarraymulticol{3}{l}{\I(X_1;Y_1|H_1)}\nonumber\\
  \quad & = & \h(H_1X_1+Z_1|H_1)-\h(Z_1) \nonumber\\
  & = & \int_{|h_1| \geq \Gamma} \h(H_1X_1+Z_1|H_1=h_1) \d F_{H_1}(h_1) \nonumber\\
  & & {} +\int_{|h_1| < \Gamma} \h(H_1X_1+Z_1|H_1=h_1) \d
  F_{H_1}(h_1) - \h(Z_1) \nonumber\\
  & \geq & \int_{|h_1| \geq \Gamma} \h(H_1X_1+Z_1|H_1=h_1) \d
  F_{H_1}(h_1) \nonumber\\
  & & {} + \Prv{|H_1|<\Gamma} \h(Z_1)-\h(Z_1) \nonumber\\
  & \geq & \int_{|h_1|\geq \Gamma}\left(\log
  |h_1|^2+\h(X_1)\right)\d F_{H_1}(h_1) \nonumber\\
  &  & {} +\Prv{|H_1|<\Gamma} \h(Z_1)-\h(Z_1) \nonumber\\
  & \geq & \Prv{|H_1| \geq \Gamma}\left(\log\Gamma^2+\h(X_1)\right)\nonumber\\
  && {} + \Prv{|H_1|<\Gamma} \h(Z_1) - \h(Z_1) \nonumber\\
  & = & \Prv{|H_1| \geq
    \Gamma}\left(\log\Gamma^2+\log\pi+\h(|X_1|^2)\right) \nonumber\\
  & & {} + \Prv{|H_1|<\Gamma} \h(Z_1) - \h(Z_1) \nonumber\\
  & = & \Prv{|H_1| \geq \Gamma}\log \const{A}^2 +
  \Prv{|H_1|\geq\Gamma}\log\left(\pi\Gamma^2\right) \nonumber\\
  && {} + \Prv{|H_1|<\Gamma}
  \h(Z_1) - \h(Z_1)\nonumber\\
  & = & \Prv{|H_1| \geq \Gamma}\log \const{A}^2 +
  \Prv{|H_1|\geq\Gamma}\log\left(\pi\Gamma^2\right) \nonumber\\
  && {} +
  \left(\Prv{|H_1|<\Gamma}-1\right)\log(\pi e\sigma^2) \nonumber\\
  & = &  \Prv{|H_1| \geq \Gamma}\log \SNR -
  \Prv{|H_1|\geq\Gamma}\left(1-\log\Gamma^2\right)
  \label{eq:coherent}
\end{IEEEeqnarray}
where $F_{H_1}(\cdot)$ denotes the distribution function of the fading $H_1$.
Here, the first inequality follows by conditioning on $X_1$; the
second by conditioning on $Z_1$ and by the behavior of differential
entropy under scaling; the next inequality because over the
range of integration $|h_1|\geq \Gamma$ we have $\log|h_1|^2 \geq
\log\Gamma^2$; the subsequent equality follows because $X_1$ is circularly-symmetric \cite[Lemma
6.16]{lapidothmoser03_3}; and the next one follows by computing the entropy
of a random variable that is uniformly distributed on the interval
$\left[0,\const{A}^2\right]$.

We next turn to the second term on the RHS of \eqref{eq:1}. In order
to upper bound it we proceed along the lines of
\cite{denghaimovich04}, but for non-Gaussian fading. Let $\vect{Y}$, $\vect{H}$, and $\vect{Z}$ be
the respective random vectors $\trans{(Y_1,\ldots,Y_n)}$,
$\trans{(H_1,\ldots,H_n)}$, and $\trans{(Z_1,\ldots,Z_n)}$ with
$\trans{\vect{A}}$ denoting the transpose of $\vect{A}$. With this
notation, \eqref{eq:channel} is equivalent to
\begin{equation}
  \vect{Y} = \mat{X} \vect{H} + \vect{Z}
\end{equation}
where $\mat{X}$ is a diagonal matrix with diagonal entries $x_1,\ldots,x_n$. It follows that the covariance matrix of $\vect{Y}$
given $x_1,\ldots,x_n$ can be written as
\begin{IEEEeqnarray}{lCl}
  \IEEEeqnarraymulticol{3}{l}{\Econd{\left(\vect{Y}-\E{\vect{Y}}\right)\hermi{\left(\vect{Y}-\E{\vect{Y}}\right)}}{X_1=x_1,\ldots,X_n=x_n}}\nonumber\\
  \IEEEeqnarraymulticol{3}{r}{= \mat{X}
  \mat{K}_{\vect{H}\vect{H}}\hermi{\mat{X}}
  + \sigma^2 \mat{I}_n \quad}
\end{IEEEeqnarray}
where $\mat{I}_n$ is the $n \times n$ identity matrix,
$\hermi{(\cdot)}$ denotes Hermitian conjugation, and 
\begin{equation}
  \mat{K}_{\vect{H}\vect{H}} \triangleq \E{(\vect{H}-\E{\vect{H}})\hermi{(\vect{H}-\E{\vect{H}})}}.
\end{equation}
Using the entropy maximizing
property of circularly-symmetric Gaussian vectors \cite[Theorem 9.6.5]{coverthomas91}, we have
\begin{IEEEeqnarray}{lCl}
  \IEEEeqnarraymulticol{3}{l}{\I(H_1^n;Y_1^n|X_1^n)}\nonumber\\
  \quad & = & \h(Y_1^n|X_1^n)-\h(Z_1^n) \nonumber\\
  & \leq &
  \E{\log\det\left(\mat{I}_n+\frac{1}{\sigma^2}\vmat{X}\mat{K}_{\vect{H}\vect{H}}\hermi{\vmat{X}}\right)}\nonumber\\
  & \stackrel{\textnormal{(a)}}{=} &
  \E{\log\det\left(\mat{I}_n+\frac{1}{\sigma^2}\mat{K}_{\vect{H}\vect{H}}\hermi{\vmat{X}}\vmat{X}\right)}\nonumber\\
  & \stackrel{\textnormal{(b)}}{\leq} &
  \log\det\left(\mat{I}_n+\frac{\const{A}^2}{\sigma^2}\mat{K}_{\vect{H}\vect{H}}\right)
  \nonumber\\
  & = & \log\big((\pi e)^n
  \det\left(\mat{I}_n+\SNR\mat{K}_{\vect{H}\vect{H}}\right)\big)-\log(\pi
  e)^n\nonumber\\
  & \stackrel{\textnormal{(c)}}{=} & h(V_1,\ldots,V_n)-\log(\pi e)^n\label{eq:noncoherent}
\end{IEEEeqnarray}
where $\vmat{X}$ is a random diagonal matrix with diagonal entries $X_1,\ldots,X_n$;
and where $\{V_k\}$ is a zero-mean, stationary \& ergodic, circularly-symmetric,
complex Gaussian process whose spectral distribution function
$F_V(\cdot)$ is given by
\begin{equation}
  \label{eq:vprocess}
  F_V(\lambda) = \lambda+\SNR F(\lambda), \qquad -1/2\leq \lambda \leq 1/2.
\end{equation}
Here, (a) follows from the identity
$\det(\mat{I}_n+\mat{A}\mat{B})= \det(\mat{I}_n+\mat{B}\mat{A})$; (b)
follows from \eqref{eq:power} which implies that $\const{A}^2
\mat{I}_n - \hermi{\vmat{X}}\vmat{X}$ is positive semi-definite with
probability one; and (c) follows from the expression of the
differential entropy of a circularly-symmetric Gaussian vector and by noting that the
covariance matrix of the random vector $\trans{(V_1,\ldots,V_n)}$ is $\mat{I}_n+\SNR\mat{K}_{\vect{H}\vect{H}}$.
Dividing \eqref{eq:noncoherent} by $n$ and taking the limit as $n$
tends to infinity yields
\begin{IEEEeqnarray}{lCl}
  \IEEEeqnarraymulticol{3}{l}{\lim_{n \to \infty} \frac{1}{n}
    \I(H_1^n;Y_1^n|X_1^n)}\nonumber\\
  \quad & \leq &
  \lim_{n \to \infty} \frac{1}{n}h(V_1,\ldots,V_n)-\log (\pi e)
  \nonumber\\
  & = & \int_{-1/2}^{1/2} \log\left(1+\SNR F'(\lambda)\right) \d\lambda  \label{eq:spectrum}
\end{IEEEeqnarray}
where the equality follows from the expression of the differential
entropy rate of a Gaussian process \cite[Section 11.5]{coverthomas91}.

Equations \eqref{eq:1}, \eqref{eq:coherent2}, \eqref{eq:coherent}, and \eqref{eq:spectrum} yield
the capacity lower bound
\begin{IEEEeqnarray}{lCl}
  \IEEEeqnarraymulticol{3}{l}{\C(\SNR)}\nonumber\\
  \quad & \geq &  \Prv{|H_1| \geq \Gamma}\log \SNR - \Prv{|H_1|\geq
  \Gamma}\left(1-\log\Gamma^2\right)
  \nonumber\\ 
  & & {} - \int_{-1/2}^{1/2} \log\left(1+\SNR
  F'(\lambda)\right)\d\lambda \label{eq:LB}
\end{IEEEeqnarray}
for any fixed $\Gamma>0$. Note that this lower bound holds for all mean-$d$,
unit-variance, stationary \& ergodic fading processes $\{H_k\}$ with
spectral distribution function $F(\cdot)$.

\section{Asymptotic Analysis}
\label{sec:prelog}
In the following we prove Theorem~\ref{thm:prelog} by computing the
limiting ratio of the capacity lower bound \eqref{eq:LB} to the
logarithm of the SNR, as the SNR tends to infinity.

We first show that
\begin{IEEEeqnarray}{lCl}
  \IEEEeqnarraymulticol{3}{l}{\lim_{\SNR \to \infty} \int_{-1/2}^{1/2} \frac{\log\left(1+\SNR
        F'(\lambda)\right)}{\log\SNR}\d\lambda}\nonumber\\
  \qquad \qquad \qquad \qquad \qquad & = & \mu\left(\left\{\lambda:F'(\lambda)>0\right\}\right).\IEEEeqnarraynumspace\label{eq:all}
\end{IEEEeqnarray}
For this purpose we divide the integral into three parts, depending
on whether $\lambda$ takes part in the set $\set{S}_1$, $\set{S}_2$,
or $\set{S}_3$, where
\begin{IEEEeqnarray}{lCl}
  \set{S}_1 & \triangleq & \{\lambda \in [-1/2,1/2]:\,F'(\lambda)=0\}\\
  \set{S}_2 & \triangleq & \{\lambda \in [-1/2,1/2]:\,F'(\lambda)\geq
  1\}\\
  \set{S}_3 & \triangleq & \{\lambda \in [-1/2,1/2]:\,0<F'(\lambda)<1\}.
\end{IEEEeqnarray}
 For $\lambda \in \set{S}_1$ the integrand is zero and hence
\begin{equation}
  \lim_{\SNR\to\infty} \int_{\set{S}_1}\frac{\log\left(1+\SNR
      F'(\lambda)\right)}{\log\SNR}\d\lambda = 0.\label{eq:zero}
\end{equation}
For $\lambda \in \set{S}_2$ so that $F'(\lambda) \geq 1$ we note that for sufficiently large SNRs the
function
\begin{equation*}
  \frac{\log\left(1+\SNR
      F'(\lambda)\right)}{\log\SNR}
\end{equation*}
is monotonically decreasing in the SNR. Therefore, applying the
Monotone Convergence Theorem, we have
\begin{IEEEeqnarray}{lCl}
  \IEEEeqnarraymulticol{3}{l}{\lim_{\SNR\to\infty}
    \int_{\set{S}_2}\frac{\log\left(1+\SNR
        F'(\lambda)\right)}{\log\SNR}\d\lambda}\nonumber\\
  \quad & = &  \int_{\set{S}_2} \lim_{\SNR\to\infty}\frac{\log\left(1+\SNR
      F'(\lambda)\right)}{\log\SNR}\d\lambda \nonumber\\
  & = & \mu\left(\set{S}_2\right)\nonumber\\
  & = & \mu\left(\left\{\lambda:F'(\lambda)\geq 1\right\}\right).\label{eq:monotone}
\end{IEEEeqnarray}
For $\lambda \in \set{S}_3$ so that $0<F'(\lambda)<1$ we have
\begin{IEEEeqnarray}{lCl}
  \IEEEeqnarraymulticol{3}{l}{0 \leq \frac{\log\left(1+\SNR
      F'(\lambda)\right)}{\log\SNR} < \frac{\log(1+\SNR)}{\log\SNR} \leq
      \log(1+e),}\nonumber\\
    \IEEEeqnarraymulticol{3}{r}{\quad \SNR \geq e,\,\,\,\,\,}
\end{IEEEeqnarray}
where the last inequality follows because for sufficiently large
SNRs the function $\log(1+\SNR)/\log\SNR$ is monotonically
decreasing in the SNR. Since $\log(1+e)$ is integrable over $\set{S}_3$ we can make
use of the Dominated Convergence Theorem to obtain
\begin{IEEEeqnarray}{lCl}
  \IEEEeqnarraymulticol{3}{l}{\lim_{\SNR\to\infty}
    \int_{\set{S}_3}\frac{\log\left(1+\SNR
        F'(\lambda)\right)}{\log\SNR}\d\lambda}\nonumber\\
  \quad & = & \int_{\set{S}_3} \lim_{\SNR\to\infty}\frac{\log\left(1+\SNR
      F'(\lambda)\right)}{\log\SNR}\d\lambda \nonumber\\
  & = & \mu\left(\set{S}_3\right) \nonumber\\
  & = & \mu\left(\left\{\lambda:0<F'(\lambda)<1\right\}\right).\label{eq:dominated}
\end{IEEEeqnarray}
Adding \eqref{eq:zero}, \eqref{eq:monotone}, and
\eqref{eq:dominated} yields \eqref{eq:all}.

To continue with the asymptotic analysis of \eqref{eq:LB} we now note
that by \eqref{eq:all}
\begin{IEEEeqnarray}{lCl}
  \Pi & \triangleq &
  \varlimsup_{\SNR\to\infty}\frac{\C(\SNR)}{\log\SNR}\nonumber\\
  & \geq & \Prv{|H_1|\geq \Gamma} -
  \mu\left(\left\{\lambda:F'(\lambda)>0\right\}\right)\nonumber\\
  & = & \mu\left(\left\{\lambda:F'(\lambda)=0\right\}\right)-\Prv{|H_1|<\Gamma}\label{eq:withgamma}
\end{IEEEeqnarray}
for any $\Gamma>0$. Since, by the theorem's assumption,
$\Prv{H_1=0}=0$, the cumulative distribution of $|H_1|$ is continuous
at zero so that
\begin{equation}
  \lim_{\Gamma \downarrow 0}\Prv{|H_1|<\Gamma} =0
\end{equation}
and Theorem~\ref{thm:prelog} therefore follows from \eqref{eq:withgamma} by
letting $\Gamma$ tend to zero from above.

\section{Summary and Conclusion}
\label{sec:discussion}
In this paper we showed that, among all stationary \& ergodic fading
processes $\{H_k\}$ with spectral distribution function $F(\cdot)$ and
satisfying \eqref{eq:condition}, the Gaussian process gives rise to
the smallest capacity pre-log. This demonstrates the robustness of the
Gaussian assumption in the analysis of fading channels at high SNR.



The result can be extended easily to multiple-input
single-output (MISO) fading channels with memory when the fading processes
corresponding to the different transmit antennas are independent of
each other. An expression for the capacity pre-log for MISO Gaussian
fading can be found in \cite{kochlapidoth05_1},
\cite{kochlapidoth05_3}.


\end{document}